\documentclass[%
 reprint,
 amsmath,amssymb,
 aps,
prl,
]{revtex4-2}

\usepackage{amssymb}
\usepackage{color}
\usepackage{graphicx}
\usepackage{dcolumn}
\usepackage{bm}

\usepackage{url}
\usepackage{hyperref} 

\usepackage{amsmath}
\hypersetup{
    colorlinks=true,      
    linkcolor=blue,       
    filecolor=magenta,    
    urlcolor=blue,        
    citecolor=blue,       
}

\begin{document}

\preprint{APS/123-QED}

\title{Spatially Indirect Exciton Condensation \\ in Two-Dimensional Strongly Correlated Semimetals}

\author{Yao Zeng}
\affiliation{Guangdong Provincial Key Laboratory of Magnetoelectric Physics and Devices}
\affiliation{ School of Physics, Sun Yat-sen University, Guangzhou, Guangdong 510275, China
}

\author{Shi-Cong Mo}
\affiliation{Guangdong Provincial Key Laboratory of Magnetoelectric Physics and Devices}
\affiliation{ School of Physics, Sun Yat-sen University, Guangzhou, Guangdong 510275, China
}

\author{Xiang Chen}
\affiliation{Guangdong Provincial Key Laboratory of Magnetoelectric Physics and Devices}
\affiliation{ School of Physics, Sun Yat-sen University, Guangzhou, Guangdong 510275, China
}

\author{W\'ei W\'u}
\email[Corresponding author: ]{wuwei69@mail.sysu.edu.cn}
\affiliation{Guangdong Provincial Key Laboratory of Magnetoelectric Physics and Devices}
\affiliation{ School of Physics, Sun Yat-sen University, Guangzhou, Guangdong 510275, China
}

\date{\today}


\begin{abstract}
Identifying materials hosting an excitonic insulator ground state has been one of the major pursuits in condensed matter physics in recent years.
 Promising candidates in transition metal chalcogenide compounds (TMC), including $1T$-$\mathrm{TiSe_2}$, $\mathrm{Ta_2Pd_3Te_5}$, and $\mathrm{Ta_2NiSe_5}$, share a crucial common characteristic:  their low-energy physics is governed by electrons in $d$-orbitals subject to strong on-site Coulomb interactions. 
 In this work, we investigate spatially indirect exciton condensation in two-dimensional semimetals on triangular lattice. Using a combination of dynamical mean-field theory and the determinant  quantum Monte Carlo method, we study two- and three-orbital Hubbard models incorporating strong on-site  ($U$) and inter-orbital interactions ($V$). Our results demonstrate that  on-site Hubbard $U$ can strongly suppress the condensation temperature $T_c$,  an effect that is particularly pronounced at higher electron-hole pair densities.  This behavior contrasts sharply with the case  without on-site $U$, where $T_c$ grows with pair density at fixed $V$.
 Moreover, we uncover competition among multiple electron-hole pairing channels in the three-orbital model, which also acts to suppress $T_c$ of exciton condensation. An orbital-selective electron-hole pairing state is identified.
These findings may help explain the large discrepancy between strong binding-energy and relative
low transition temperature for  indirect excitons in TMCs materials, offering important insights for 
understanding and engineering exciton condensation in materials with strongly correlated $d$- shell electrons.

\end{abstract}

\keywords{Excitonic Insulator, Multi-orbital Hubbard Model, Strongly Correlated Systems}

\maketitle

\textit{Introduction -}
The excitonic insulator (EI) constitutes a paradigmatic correlated state arising from a bosonic condensation of Coulomb-bound electron-hole pairs in narrow-gap semiconductors or semimetals~\cite{mott1961transition,blatt1962bose,jerome1967excitonic, eisenstein2004bose}. This fundamental phase of matter has been investigated extensively in both experiment~\cite{cercellier,jiang2019,ma2021strongly,baldini2023spontaneous,zhang2024spontaneous,huang2024,hossain2025topological,gao2024observation,wang2026resonance,meng2025isolating,liu2025probing} and theory~\cite{blatt1962bose,jerome1967excitonic,lozovik1976new,wu2015,jiang17,jiang2019,guan2023,shao2024,yao2024excitonic,zeng2025topological,xia2023full}. The spatially separated electron--hole bilayer structures~\cite{eisenstein2004bose} provide particularly favorable platforms for achieving this state, as spatial separation suppresses electron-hole recombination, hence prolonging exciton lifetimes and promotes the condensation of indirect excitons. The exploration of many exotic emergent phenomena related to EI, such as the BCS-BEC crossover~\cite{sreejith2024eliashberg,giuli2023mott,zhu2010exciton,zhu2024interaction,lopez2018evidence,nilsson2021effects,moon2025exciton,gao2023evidence,wang2019evidence},
supersolidity~\cite{conti23,zhang21,dai24},  finite momentum pairing 
~\cite{chen1991excitonic,bi2021excitonic,zeng2023exciton,dong25}, excitonic topological phases~\cite{wang2023excitonic,hossain2025topological}, and perfect
Coulomb drag\cite{nguyen2025perfect,qi2025perfect}, can thus be facilitated in the indirect exciton systems.

Quantum wells in $\mathrm{InAs}/\mathrm{GaSb}$ heterostructures have served as pioneering platforms in the pursuit of these exotic phases~\cite{zhu95,wang2023excitonic}. These semiconductor systems revealed signatures of excitonic topological order and superfluidity~\cite{wang2023excitonic,snoke2002spontaneous,eisenstein2004bose}.
More recently, transition metal chalcogenide (TMC) platforms, such as $\mathrm{MoSe}_2/\mathrm{hBN}/\mathrm{WSe}_2$ heterostructures, have been pivotal systems for exploring high-temperature EI~\cite{nguyen2025perfect,qi2025perfect}.
 By exploiting Type-II band alignment to spatially separate carriers, these devices host long-lived excitons that exhibit macroscopic coherence, evidenced by nonlinear electroluminescence up to $100~\text{K}$~\cite{wang2019evidence}. Crucially,
  transport measurements revealed perfect Coulomb drag at zero magnetic field~\cite{nguyen2025perfect,qi2025perfect}, establishing a key foundation for realizing frictionless flow of charge-neutral excitons~\cite{ulman2021organic,regan2022emerging,nandi2012exciton,drag09} in transport experiments.
In bulk EI materials, on the other hand, the layered TMC compounds $1T$-$\mathrm{TiSe_2}$~\cite{kogar2017signatures,cercellier,ou2024incoherence,ye2026ultrafast}, $\mathrm{Ta_2Pd_3Te_5}$~\cite{huang2024,zhang2024spontaneous}, and $\mathrm{Ta_2NiSe_5}$ \cite{yao2024excitonic,yu2024observation,huang2024,zhang2024spontaneous,bae2025microscopic,PhysRevResearch.5.043089,li2024disentangling}
 have sparked intense debate regarding the existence of exciton condensation.
This is because the gap opening in these materials, a hallmark of the excitonic ordered state, can also be attributed to non-excitonic structure phase transitions~\cite{mazza20,liu2021photoinduced,baldini2023spontaneous,wei2025gate,chen2025structural}. Compared to $\mathrm{Ta_2NiSe_5}$, which may host significant structural distortion~\cite{kim2021direct,wei2025gate}, $\mathrm{Ta_2Pd_3Te_5}$ offers a cleaner paradigm,  as in which the lattice distortion is negligible in the insulating phase~\cite{zhang2024spontaneous,huang2024}. This structural stability strongly supports an exciton instability driven by Coulomb interactions in $\mathrm{Ta_2Pd_3Te_5}$.

While experimental advances in the study of EI have progressed rapidly, a comprehensive theoretical understanding~\cite{kozlov1965metal,bronold2006possibility,fogler2014high,wu2015,combescot2017bose,sreejith2024eliashberg,dai24,kaneko2025new} of this phase in real materials remains incomplete. We notice that in theory, large exciton binding energies being order of $\sim \mathrm{1eV}$ have been predicted in semi-metals or narrow gap semiconductors~\cite{yao2024excitonic, tang2025,zhao2025one,sun2025breaking}, a high temperature EI  in experiments,  however, is still elusive. 
Conventional theoretical frameworks are often established on the basis of mean-field or perturbative approaches. In EI candidate materials based on TMC, the strong on-site Hubbard interaction $U$ in $d$-orbitals is usually larger than the electron bandwidth $W$~\cite{ramezani2024nonconventional},  and it can significantly exceed the inter-orbital (inter-layer) electron-hole attraction $V$~\cite{mazza20}. This strong-correlation regime  calls into question the adequacy of weak-coupling theories that consider $V$ while neglecting the substantial on-site interaction $U$. 

In this work, we investigate excitonic condensation in two-dimensional (2D) semimetals using a combination of Cellular dynamical Mean-Field theory (CDMFT)~\cite{RevModPhys.68.13} and determinant quantum Monte Carlo (DQMC)~\cite{blankenbecler1981monte}. The non-perturbative CDMFT approach enables us to capture the quantum fluctuation of strongly interacting $d$- electrons beyond weak-coupling treatments. We show that the exciton condensation temperature $T_c$ can be significantly suppressed by on-site Coulomb repulsion $U$, although it in general remains non-zero for finite $V$.
We also show that a finite $U$ fundamentally modifies the density dependence of $T_c$. Specifically, $T_c$ drops with increasing carrier concentration $n$, which contrasts with the $U=0$ case, where $T_c$ grows with pair density $n$ at fixed $V$. Using a three-orbital model, we further uncover the competition between different pairing channels, which provides an additional mechanism suppressing $T_c$. An orbital-selective EI state is also revealed in this model.
Finally, we discuss the implications of these findings for experimental observations.

\begin{figure}[t!]
\centering
\includegraphics[width=0.48\textwidth]{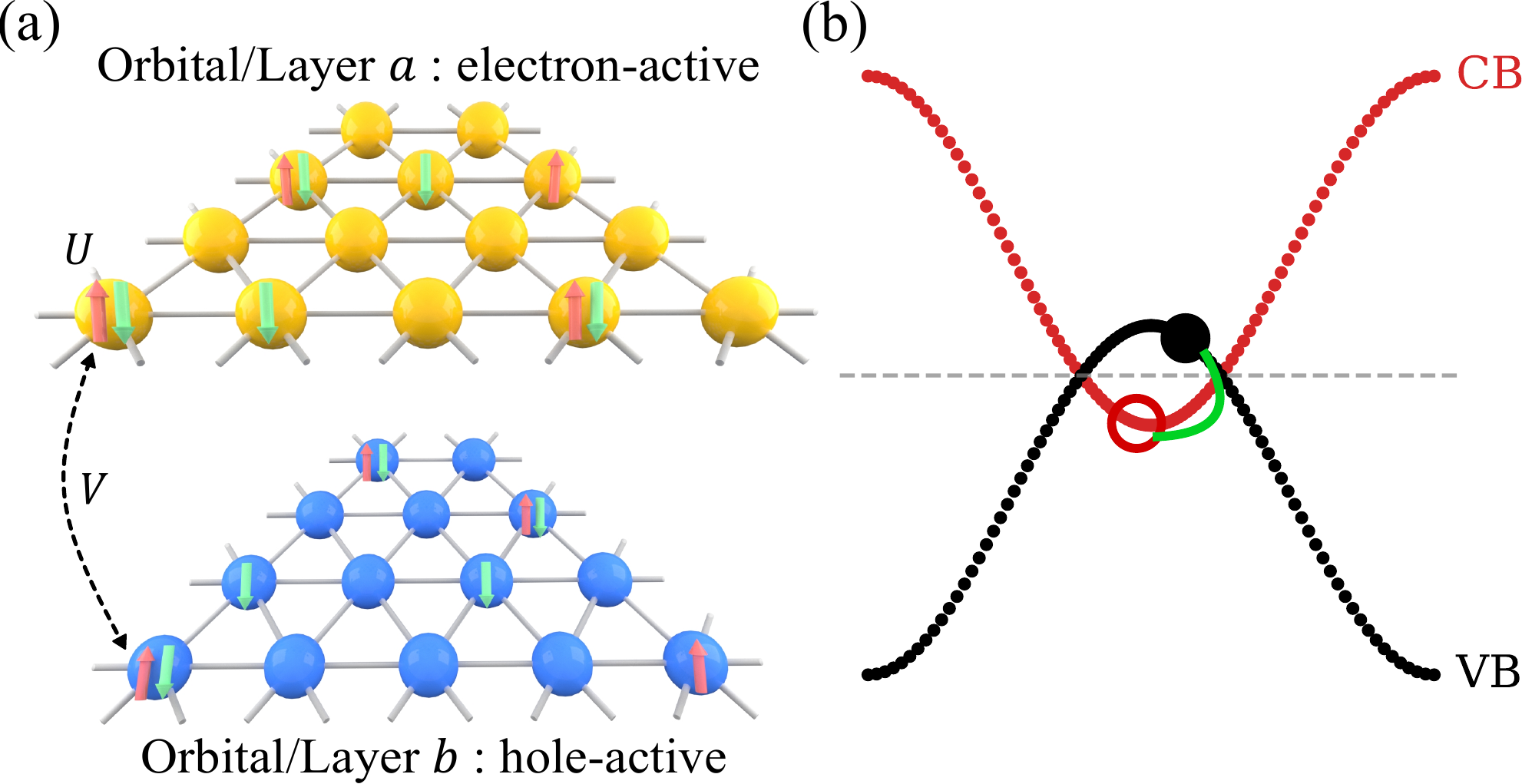}
\caption{\label{fig1}\textbf{(a)} Schematic illustration of the two-orbital (two-layer) Hubbard model on 2D triangular lattice.
$U$ represents the  on-site intra-orbital(intra-layer) Coulomb interaction, while $V$ denotes the inter-orbital (inter-layer) Coulomb interaction in a unit cell. 
The orbital/layer $a$ (upper) is designated as electron-active, while orbital/layer $b$ (lower) acts as hole-active.
Dotted line indicates a possible electron--hole pairing.
\textbf{(b)} Schematic illustration of the non-interacting band structure we use which is a semimetal at $U=0, V=0$.
Upper/lower curves indicate the conduction/valence bands (CB/VB).
The binding of holes at the CB bottom (hollow circles) and electrons at the VB top (filled circles) leads to the formation of indirect excitons (indicated by the solid line between the two circles). Dashed line indicates the Fermi level.}
\end{figure}


\textit{Model and Methods -}
Our two-orbital Hubbard model we use is given by,
\begin{align}
H &= -\sum_{\langle ij \rangle \sigma m} t_m c^{\dagger}_{i\sigma m} c_{j\sigma m} 
     + \text{H.c.} -  \sum_{i\sigma m} (\mu - \epsilon_m) \hat{n}_{i\sigma m} \notag \\
  &\quad +   \sum_{im} U_m \hat{n}_{i\uparrow m} \hat{n}_{i\downarrow m} 
     + V \sum_{i\sigma\sigma^{\prime}, m \neq m^{\prime}} \hat{n}_{i\sigma m} \hat{n}_{i\sigma^{\prime}m^{\prime}}
\end{align}
where the $m=a,b$ labels the orbital (or layer) degrees of freedom, corresponding to orbitals in bulk systems and layers in 2D materials, respectively. 
 $c_{i\sigma m}$ ($c^{\dagger}_{i\sigma m}$) denotes the electron annihilation (creation) operator at unit cell $i$, orbital (layer) $m$, with spin $\sigma$; $n_{i\sigma m}$ is the particle number operator. The nearest-neighbor hopping integral $t_a$ is set to $t_a=t=1$ as energy unit throughout the paper. $U_m$ is the on-site repulsion between electrons at orbital (layer) -$m$, $V$ is the local inter-orbital (inter-layer) interaction. In this work, we have omitted long-range Coulomb interactions to simplify the numerical study.  
 We adjust chemical potential $\mu$ and orbital(layer) energy $\epsilon_m$ to tune the orbital (layer) densities.
 Here we adopt a pure electron picture to study the problem, \textit{i.e.}, $\langle \hat{n_a} \rangle = n_e \equiv n$ denote the carrier  density in  the electron orbital (orbital-$a$), implying a hole orbital (layer) of $ n_h = 1- \langle \hat{n_b} \rangle =n_e = n$ in the orbital-$b$. We assume equal mass for both carriers, \textit{i.e.}, $t_a = -t_b$.
 The schematic of our model and the simplified semimental band structure are illustrated in Fig.~\ref{fig1}. Orbitals (layers) $a$ (upper) and $b$ (lower) are designated as electron-active and hole-active, respectively.
 Previous studies have used the two-orbital Hubbard model to investigate excitonic effects in the strongly correlated regime~\cite{Kunes_2015}, such as the transition between  excitonic insulator state and  antiferromagnetic state~\cite{PhysRevB.85.165135}, as well as the interplay between the excitonic insulator and low-/high-spin states~\cite{PhysRevB.89.115134}.
 
Our theoretical investigation of the bilayer Hubbard model on a triangular lattice is carried out by the Cellular Dynamical Mean-Field Theory, a cluster extension of CDMFT ~\cite{RevModPhys.68.13} designed to incorporate non-local spatial correlations.
 In this framework, the lattice problem is mapped onto an effective quantum impurity model for a small cluster, which is then solved self-consistently within a dynamical bath
  that represents the rest of the lattice degrees of freedom. Here we typically use a two-site effective cluster incorporating $a,b$ orbitals within a unit cell. 
  The effective cluster  problem is solved using the continuous-time quantum Monte Carlo method (CTQMC)\cite{SETH2016274,Parcollet_2015}. 
  Our simulations typically accumulate $10^9$  Monte Carlo sweeps in each CDMFT loop. Additional details on the numerical method can be found in the Supplementary Information.

\begin{figure}[t!]
\centering
\includegraphics[width=0.4\textwidth]{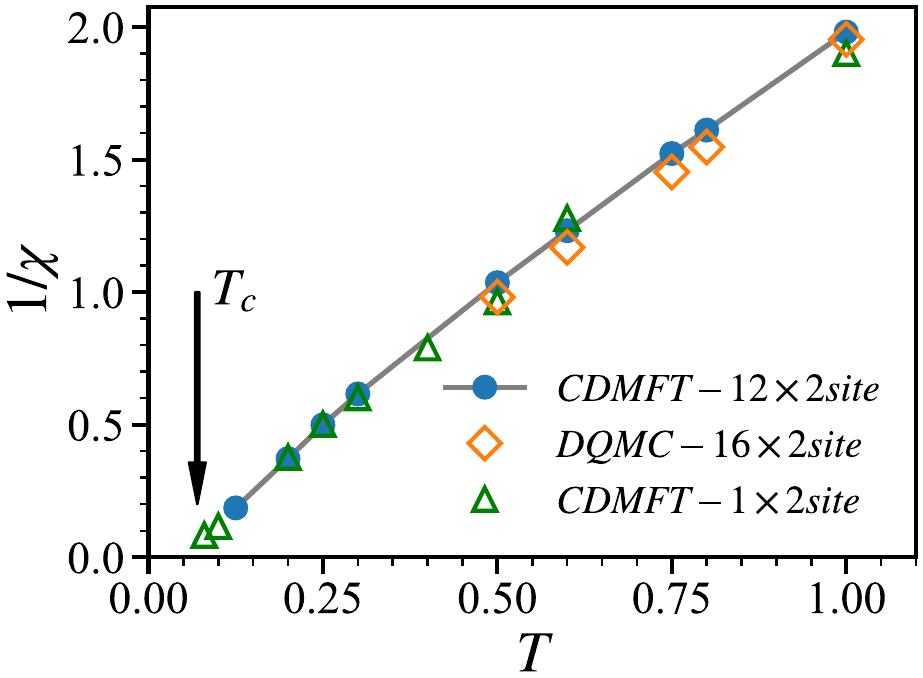}
\caption{\label{fig2} The inverse susceptibility $1/\chi$ as a function of temperature $T$.
 When $1/\chi$ approaches zero, the system enters the excitonic condensate phase,
 and the corresponding temperature is defined as the critical transition temperature $T_c$. Here, $U_a=U_b=1,V=4$, density, $\langle n_a \rangle= 0.2, \langle n_b \rangle = 0.8$ (thus $n_e = n_h = 0.2$). An excitonic condensation $T_c$ is found about $ T_c \sim 0.07$.
 The DQMC calculations were performed on a $4 \times 4 \times 2$ lattice, while CDMFT calculations were performed on $1 \times 2$- and $12 \times 2$-site clusters.}
\end{figure}

\textit{Result -}
The onset of the excitonic condensation phase can be identified by the divergence of the uniform excitonic pairing susceptibility $\chi$, which can be defined as,
\begin{eqnarray}
\chi=\frac{1}{N}\sum_{i,j}\int_{0}^{\beta}d\tau\langle\mathcal{T}_{\tau} p_{j}(\tau)p_{i}(0)\rangle
\end{eqnarray}
where $p_i$ is the uniform spin-singlet electron-hole pair operator~\cite{giuli2023mott}, $p_i = \frac{1}{\sqrt{2}}\sum_{\sigma} \langle c^{\dagger}_{ia\sigma} c_{ib\sigma} \rangle$, and $N$ is the number of unit cells in the system.
Thus the vanishing inverse pairing susceptibility $1/\chi \to 0$ can be used to determine the critical transition temperature $T_c$ for excitonic condensation ( \textit{i.e.}, $T_c$ for EI).
We note that in our study, spin-singlet and spin-triplet excitons are degenerated in energy due to the symmetry of our Hamiltonian in Eq.1. In real materials, the energy difference between singlet and triplet interlayer excitons may depend on detailed material properties~\cite{wang2019giant,li2024spin,durmus2024quantum}.
In Fig.~\ref{fig2}, we plot the CDMFT result of $1/\chi$ as a function of temperature  $T$ for a typical parameter set: $U_a=U_b=1$, $V=4$, and electron-hole density $  n  =0.2$.
As the temperature decreases from $T = 1.0$ to $T = 0.1$, $1/\chi$ drops  monotonically from $1/\chi \approx 2.0$ to $1/\chi \approx 0.1$, signaling a growing tendency to El condensation as $T$ reduces. As $T$ approaches $T_c \approx 0.07$, $1/\chi$ vanishes, denoting an El instability in the system at the given parameters.
We observe excellent agreement of $1 \times 2$ - and $12 \times 2$ - site CDMFT with exact high-$T$ DQMC results on a $4 \times 4 \times 2$ - site lattice, thereby justifying the use of CDMFT to reliably capture the exciton physics.

\begin{figure}[b!]
\centering
\includegraphics[width=0.4\textwidth]{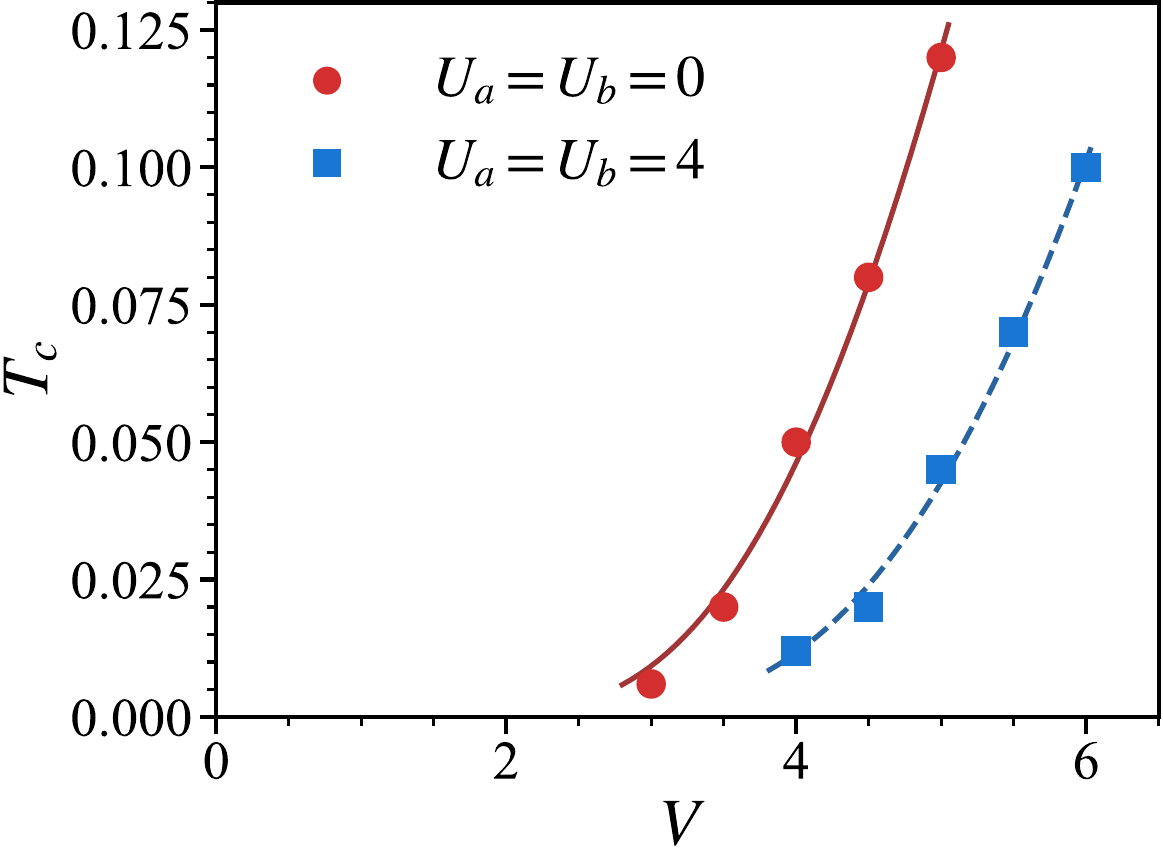}
\caption{\label{fig3}Transition temperature $T_c$ as a function of the inter-orbital interaction $V$ at different on-site interactions $U_a=U_b\equiv U$.
 Here, electron-hole density is fixed at $n_e = n_h=0.06$,  data points show results for $U=0$ (dots) and $U=4$ (squares).
 Lines are guides to the eye.}
\end{figure}

We now focus on the dependence of critical temperature $T_c$ on the inter-orbital electron-hole attraction $V$. In Fig.~\ref{fig3}, results of $T_c$ as a function of $V$ are shown, 
where calculations are performed at a fixed electron-hole density $n=0.06$, and two on-site repulsions $U=0, U=4$.
In the absence of on-site repulsion ($U=0$), $T_c$ increases from $T_c \approx 0.006$ to $T_c \approx 0.12$  as $V$ is increased from $V=3$ to $V=5$.
 For the finite $U$ case ($U=4$), $T_c$ grows from $T_c \approx 0.012$ to $T_c \approx 0.1$  as $V$ increases from $V=4$ to $V=6$. At given $V$,  
 $T_c$ is consistently smaller for $U=4$ than for $U=0$, indicating on-site $U$ plays a detrimental role in excitonic condensation.

\begin{figure}[t!]
\centering
\includegraphics[width=0.4\textwidth]{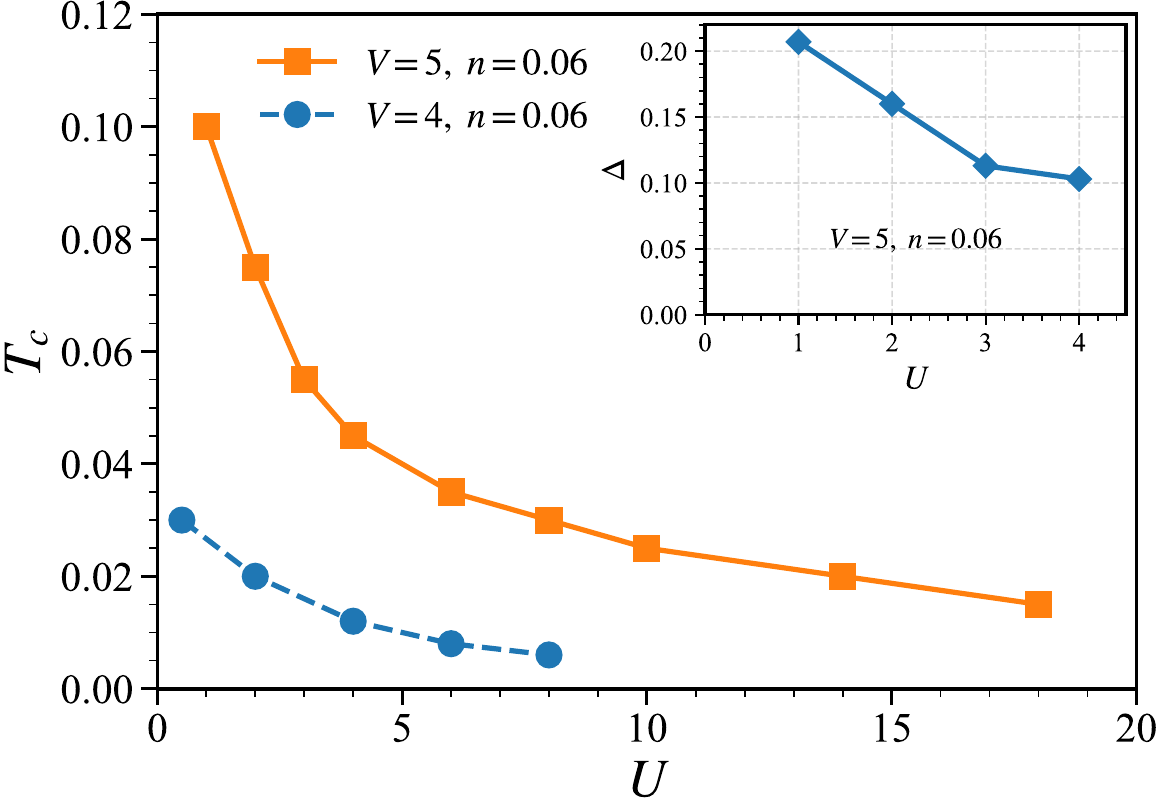}
\caption{\label{fig4}\textbf{Main panel:}Transition temperature $T_c$ as a function of the on-site Coulomb interaction $U$ at two values of inter-orbital interaction $V$. 
Squares correspond to $V=5$, while diamonds correspond to $V=4$. Here  $n_e = n_h =   n  = 0.06 $, and $1 \times 2-$ site  CDMFT is used.
\textbf{Inset}: Exciton gap $\Delta$ as functions of  $U$ at $V=5,n=0.06$. 
It should be noted that here $\Delta$ is evaluated at a small but finite temperature $T=0.01$, by using the maximum entropy analytical continuation.}
\end{figure}

To systematically investigate the effect of on-site Hubbard repulsion $U$ on excitonic condensation, we computed $T_c$ across a wide range of $U$ values.
As shown in FIG.~\ref{fig4}, $T_c$ decreases substantially as $U$ increased from zero. This decreasing trend is particularly rapid at smaller $U$ and becomes more gradual for larger $U$, see both curves in FIG.~\ref{fig4}.
Specifically, at fixed $V = 5$, increasing $U$ from $U=1$ to $U=18$ reduces $T_c$ from $T_c = 0.1$ to $ T_c = 0.015$, a decrease by nearly a factor of seven. Similarly, for $V = 4$, raising $U$ from $U= 0.5$ to $U=12$ lowers $T_c$ from $T_c = 0.03$ to $T_c = 0.005$, corresponding to a sixfold reduction. These results confirm that the on-site repulsion $U$ is a key factor limiting the stability of the excitonic condensed phase in the system.
In the inset of Fig.~\ref{fig4}, we present the exciton gap $\Delta$ as a function of $U$ for $V=5$ and $n=0.06$. 
This gap is extracted from the single-particle gap in the spectral function $A(k,\omega)$ within the exciton condensed phase (see also the Supplementary Information). 
We observe that $\Delta$ is also strongly suppressed by $U$, following a trend similar to that of $T_c$ as a function of $U$. This finding indicates that in our study, the strong suppression of $T_c$ by $U$ primarily stems from a reduction in the exciton gap, rather than a decrease in the phase stiffness of the excitonic condensate.

\begin{figure}[t!]
\centering
\includegraphics[width=0.4\textwidth]{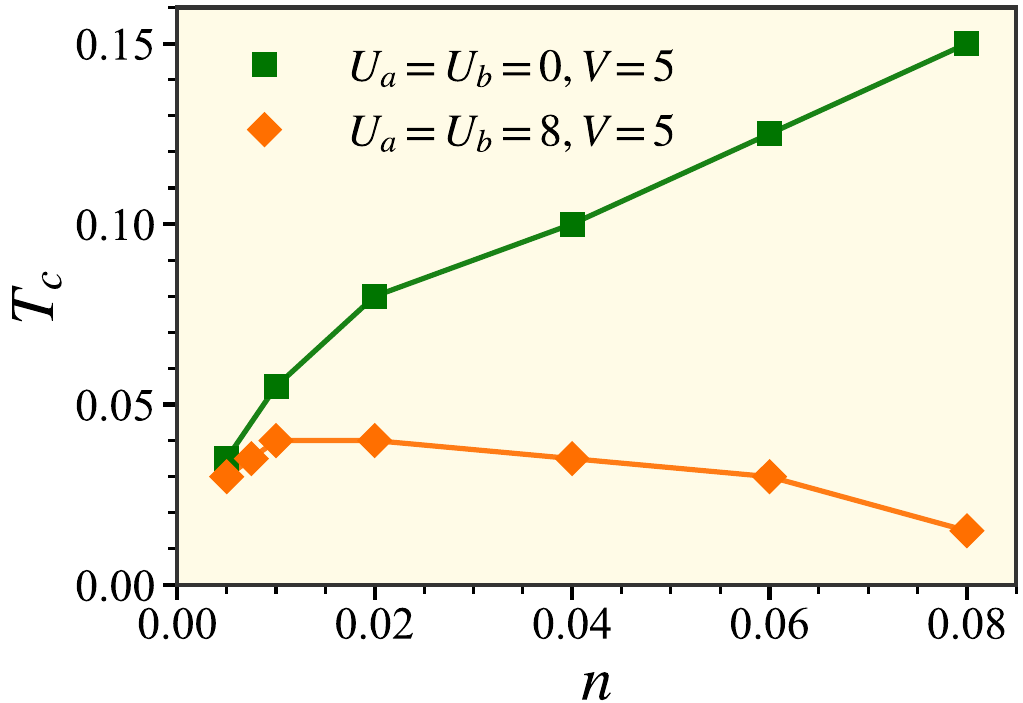}
\caption{\label{fig5}Transition temperature $T_c$ as a function of the electron--hole density at two different values of $U$. 
The inter-orbital interaction is fixed at $V=5$. $1 \times 2-$ site  CDMFT is used here.}
\end{figure}

Now we investigate the interplay between the on-site repulsion $U$ and the electron-hole density. The density dependence of $T_c$ for $U=0$  and $U=8$ is shown in FIG~\ref{fig5}.
 For the  $U=0$ case  ($U=0, V=5$), $T_c$ exhibits a monotonic increase from $T_c=0.035$ to $T_c=0.15$ as density increases from $n=0.005$ to $n=0.08$, 
 indicating that a larger population of charge carriers boosts condensation. 
In stark contrast, a finite on-site repulsion fundamentally alters this behavior. In the low-density regime ($0.005 < n < 0.01$), 
$T_c$ exhibits an initial increase for the $U=8, V=5$ case (diamonds in FIG.~\ref{fig5}) , rising from $T_c=0.03$ to $T_c=0.04$. 
As the density increases beyond $n = 0.01$, however, $T_c$ begins to decrease continuously, dropping significantly from $T_c=0.04$ to $T_c=0.015$ as $n$ increased from  $n = 0.01$ to $n=0.08$.
This non-monotonic trend at finite $U$ reflects a competition between mechanisms. At extremely low carrier density $n$, on-site repulsion $U$ plays a minor role,
 increasing $n$ enhances the coherence of electron-hole pairs, leading to higher $T_c$.
When $n$ becomes significantly large, on-site repulsion $U$ tends to dynamically bind intra-orbital local electron-hole pairs. These dynamically generated intra-orbital local electron-hole pairs have a short lifetime due to intra-orbital electron-hole recombination governed by intra-layer hopping $t_a$ and $t_b$, which prevents them from forming an intra-layer excitonic condensate. Nevertheless, they can compete with the formation of inter-orbital excitons, thereby reducing $T_c$.

In essence, these results show that in a system where low-energy physics is dominated by $d$- electrons, the formation of indirect excitons faces significant competition from intra-orbital electron-hole binding. Consequently, the system's low-energy physics is not governed purely by inter-orbital (or inter-layer) excitonic degrees of freedom. A theoretical approach that consider only inter-orbital electron-hole binding may therefore miss a key physical ingredient and severely overestimate the phase transition temperature $T_c$.
It is worthy to note that within a pure mean-field treatment (without magnetic symmetry breaking), on-site Hubbard $U$ merely introduces a Hartree shift and does not affect $T_c$ for a given electron-hole density. When further considering quasiparticle renormalization due to $U$, the dimensionless coupling constant $\tilde{\lambda}=ZVN(0)$ and effective bandwidth $\tilde{W}=ZW$ are renormalized by quasiparticle residue $Z$, where $N(0)$ is the density of states at the Fermi level. This renormalization does lead to a $U$-dependent suppression of $T_c$ in mean-field theory. However, at low densities ($n\ll0.5$) studied here, we find that $Z\sim1$. Therefore, the observed strong suppression of $T_c$ by $U$ should be attributed to dynamical effects beyond a static mean-field treatment.

Above we have studied the two-orbital electron-hole model.
In materials, carriers near Fermi level can also originate from multiple $d$-orbitals, or from a hybridization with $p$-orbitals (which possess a smaller on-site $U$). For instance, in $\mathrm{Ta_2Pd_3Te_5}$ ~\cite{yao2024excitonic}, 
hole bands acquire weight from both the $\mathrm{Pd}$-$d_{xy}$-orbital and $\mathrm{Te}$-${p_x}$-orbital. In $\mathrm{Ta_2NiSe_5}$, multiple $\mathrm{Ta}$ and $\mathrm{Ni}$ $d$-orbitals contribute to the relevant electron and hole bands~\cite{ma2022}. 
 This multiplicity of active orbitals can potentially give rise to several  distinct channels for electron-hole pair condensation~\cite{mazza20}, which may compete with one another.
To explore the interplay between multiple pairing channels involving $d$- or $p$-orbital degrees of freedom, we extend our analysis to a minimal three-orbital model  \cite{cappelluti2013tight,zhu2011giant,liu2013three}.
 Without loss of generality, we consider a system comprising one electron-type orbital (orbital- $a$) and two hole-type orbitals (orbital- $b$,$c$).
In this system, an electron can, in principle, form a bound pair with a hole from either of the other two orbitals, defining two distinct pairing channels ($ab$-, or $ac$- channel).
 
\begin{figure}[t!]
\centering
\includegraphics[width=0.4\textwidth]{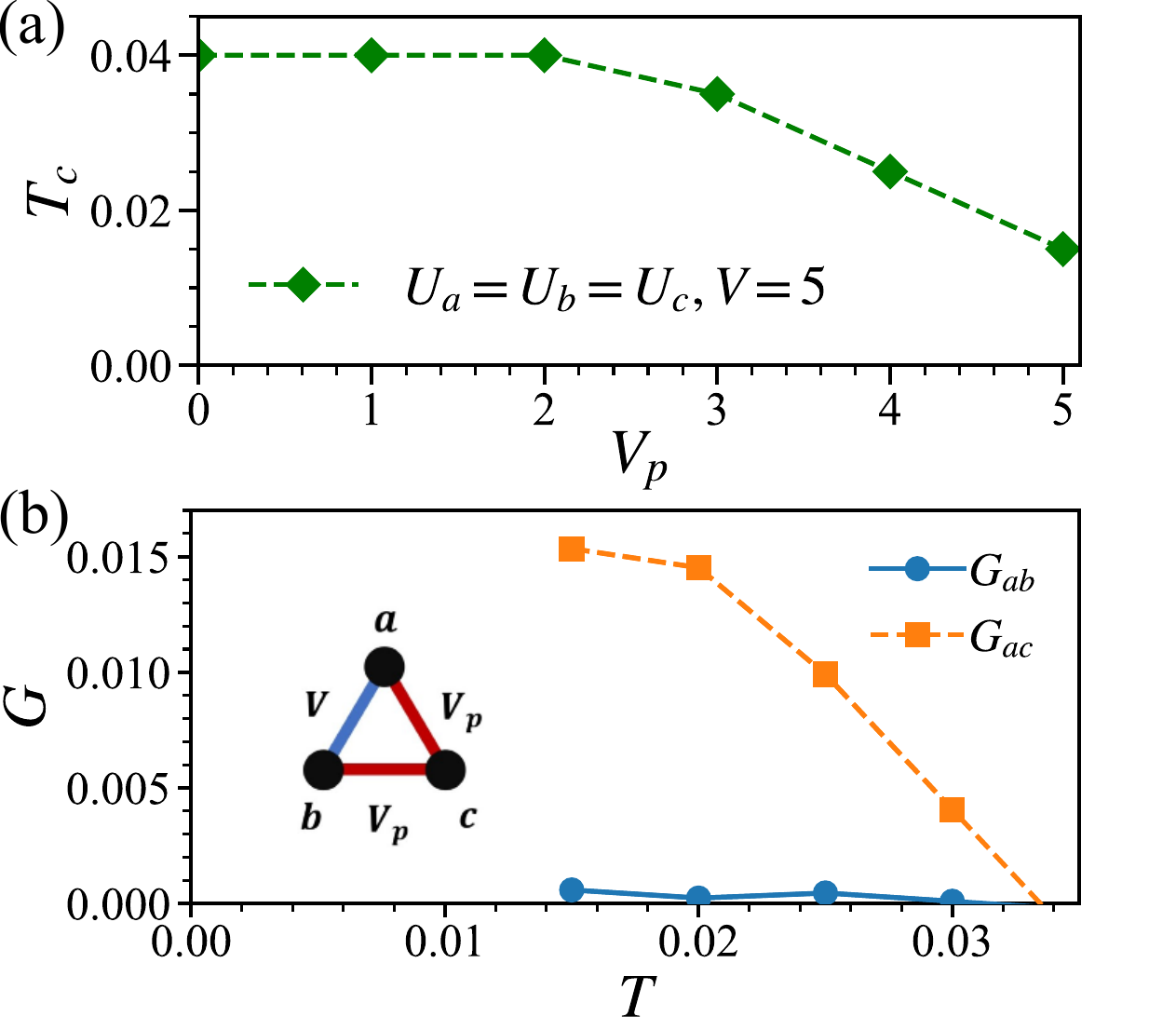}
\caption{\label{fig6}
\textbf{(a)} Transition temperature $T_c$ as a function of inter-orbital $d$-$d$ Coulomb interaction $V_{p}$.  Here  $n=0.06$,  $U_a=U_b=U_c=4$, and  $V=5$. 
\textbf{(b) Main panel}: Temperature dependence of the inter-orbital excitonic condensation order parameter $G_{ab}$ (dots) and $G_{ac}$ (squares) in the three-orbital model. Lines are guides for the eye.
\textbf{Inset} (left): Schematic of the three-orbital model, where the $a$, $b$, $c$ are distinct orbital indices. The additional orbital $c$ represents the $p$-orbital, while $a$ and $b$ represent the $d$-orbitals. 
The inter-orbital interaction between orbital $c$ and orbitals $a$ and $b$ is denoted by $V_{p}$. Electrons in one orbital can pair with holes in the other two orbitals, and a competition exists between the two types of pairing.
Here, $U_a=U_b=4,U_c=2,V=V_{p}=5$.   $1\times 2 -$ site CDMFT is used here.}
\end{figure}

 We first examine the case where all three orbitals are of $d$- electrons, thus large on-site $U$ for all three orbitals shall be assumed. Here the influence  of the additional $d$- hole orbital (orbital-$c$) is activated by gradually increasing its inter-orbital interaction strength, defined as $V_{p} \equiv V_{ac} =V_{bc}$ from zero. Simply speaking, when $V_{p} = 0$, 
 the additional hole orbital-c is fully decoupled from the system, and when $V_{p} = V_{ab}$, both hole orbitals are equally coupled to the electron orbital. The corresponding evolution of the critical temperature $T_c$ as a function of $V_{p}$ is shown
in Fig.~\ref{fig6}(a), where $U_a = U_b = U_c = 4$, and $V_{ab} =5 $ is fixed. As $V_{p}$ increased from zero,  $T_c$ initially remains nearly constant for $ V_{p} \lesssim 2$, forming a plateau. Further increasing $V_{p}$ then leads to significant suppression of $T_c$. At $V_{p} = 5$, namely, when inter-orbital interactions between three orbitals are equal, $V_{ab} = V_{bc} = V_{ac}$, the critical temperature is reduced to $T_c \approx 0.015$, a value slightly exceeding one-third of that at $V_{p} =0$. This strong suppression of $T_c$  upon introducing an additional pairing channel indicates that the competition between pairing channels, \textit{i.e.}, pairing between $ab$-, or $ac$- channels, plays a crucial detrimental role in the spontaneous formation of long-range excitonic order. 
This effect bears a conceptual resemblance to geometric frustration in magnetic systems, where competing magnetic couplings can prevent the establishment of long-range magnetic order, as seen in quantum spin liquids~\cite{zhou2017quantum}. Here, an analogous ``frustration'' arises from the presence of multiple, near-degenerate electron-hole pairing channels, which suppresses the excitonic condensation. A key distinction, however, is that in the present case, long-range excitonic order persists even at the maximally frustrated point ($V_{ab} = V_{ac}= V_{bc}$),  albeit with a significantly reduced $T_c$.
 
We then investigate the scenario with two $d$-orbitals and one $p$-orbital: a ``strong $U$'' electron orbital (orbital-$a$) and a ``strong $U$'' hole orbital (orbital-$b$) represent the $d$-orbitals, while a ``weak $U$'' hole orbital (orbital-$c$) represents the $p$-orbital. For simplicity, all inter-orbital interactions $V$ are set equal, $V_{ab}= V_{ac} = V_{bc}$.
 Fig.~\ref{fig6}(b) presents the temperature dependence of the excitonic condensation order parameters $G_{mn} =  \sum_{\sigma}\langle c^{\dagger}_{im\sigma} c_{in\sigma} \rangle $, for $(m,n) = (a,b),(a,c)$, for parameters $U_a = U_b = 4$, $U_c = 2$, and $V=5$ in this system. The results reveal a remarkable orbital-selective electron-hole pairing phenomenon. 
 Upon cooling, an excitonic order parameter between the $d\text{-}p$ orbitals ($G_{ac}$) emerges spontaneously below a critical temperature of $T_c \approx 0.032t$, and grows with further decreasing $T$. 
In contrast, the order parameter between two $d\text{-}d$ orbitals ($G_{ab}$) remains negligible at all temperatures,  despite the symmetric inter-orbital attraction ($V_{ab} = V_{ac} = V_{bc}$). This orbital-selective excitonic condensation that prefers a $d\text{-}p$  condensation over a $d\text{-}d$ condensation likely stems from the effect of Hubbard $U$, which hinders inter-orbital exciton formation. The $p$-orbital, with its smaller $U$, is more susceptible to inter-orbital electron-hole binding than a $d$-orbital with a larger $U$. Consequently, the competition between multiple channels is resolved in favor of $d\text{-}p$ exciton condensation, resulting in paired states between $d\text{-}p$ orbitals and unpaired states between the $d\text{-}d$ orbitals.

\textit{Discussion and conclusion -}
In exciton insulator candidate comprising transition metal atoms, such as $\mathrm{Ta_2Pd_3Te_5}$ and $\mathrm{Ta_2NiSe_5}$, electrons and holes are subject to multiple sources of Coulomb attractions. These multiple electron-hole binding routes effectively introduce a ``frustration'' effect that hinders the formation of exciton condensation.  Notably, the on-site Hubbard interaction $U$, typically the largest energy scale in such systems, can facilitate the formation of local electron-hole pairs or local moments, thereby substantially suppressing the condensation temperature $T_c$. While our calculations are performed for semimetallic systems with finite carrier densities in the conduction and valence bands, our conclusions should, in principle, remain valid in the semiconductor limit. In actual semiconductors, 
although the valence band is completely filled and the conduction band empty, orbital hybridization~\cite{yao2024excitonic},
 from the local atomic orbital perspective, inevitably generates holes in the valence orbit and electrons in the conduction orbital,  ensuring that the suppression of $T_c$ by on-site Hubbard $U$ persists.

 In connection with experiments, we note that for TMC excitonic insulator candidates such as $\mathrm{Ta_2Pd_3Te_5}$ and $\mathrm{Ta_2NiSe_5}$, the typical hopping integral is estimated to be $t \approx 0.1 \sim 0.2 \mathrm{eV}$~\cite{huang2024}. 
The screened on-site Hubbard repulsion $U$ for these localized $d$-orbitals typically spans $U \approx 0.8 \sim 2.5\mathrm{eV}$, while the inter-orbital interaction $V$ is on the order of $V \approx 1.0 \mathrm{eV}$~\cite{ramezani2024nonconventional,mazza20}. 
These physical energy scales yield dimensionless interaction ratios of $U/t \approx 4 \sim 25$ and $V/t \approx 5 \sim 10$, 
which fall  within our simulated parameter space. 
Based on this mapping, our calculated maximum transition temperature $T_{c} \approx 0.06t$ at typical parameters ($U=8t$, $V=6t$)
corresponds to a physical temperature of approximately $70 \sim 140 \mathrm{K}$, which is consistent with the experimentally observed $T_c \sim 100 \mathrm{K}$ in $\mathrm{Ta_2Pd_3Te_5}$~\cite{hossain2025topological}.

This agreement suggests that the strong suppression of $T_c$ by on-site Hubbard $U$, as revealed in our study, may play an important role in resolving the apparent paradox between the large exciton binding energy and the low condensation temperature observed experimentally. For example, theoretical studies have predicted a large exciton binding energy $\Delta \geq 0.6 \mathrm{eV} $~\cite{yao2024excitonic} for $\mathrm{Ta_2Pd_3Te_5}$, contrasting with its $T_c \sim 100 \mathrm{K}$, \textit{i.e.} $k_B T_c \sim 0.01 \mathrm{eV}$. 
This also offers insight into why exciton insulators remain so difficult to detect experimentally, despite the fact that exciton binding energies can be on the order of $\sim 1 \mathrm{eV}$ in low-dimensional materials~\cite{tang2025,zhao2025one,sun2025breaking}.

We propose to verify the competing mechanisms in future experiments detecting excitonic condensation temperature $T_c$ versus carrier concentration $n$, where non‑monotonic evolution may be found. On the other hand, with the polarization-dependent angle-resolved photoemission spectroscopy (ARPES)~\cite{PhysRevLett.125.216404,RevModPhys.93.025006,ma2025angle}, future experiments should be able to resolve the orbital differences of the excitonic condensation in multi-orbital excitonic insulator materials, thereby providing an approach to verify the results obtained in this work.

In summary,  we have systematically investigated spatially indirect exciton condensation in the two-dimensional Hubbard model on a triangular lattice using dynamical mean-field theory.
Our results show that the critical temperature $T_c$ for exciton condensation can be strongly suppressed by intra-orbital Coulomb repulsion $U$ of the $d$-orbitals. We uncover a remarkable contrast in the density dependence of $T_c$ between $U=0$ and $U\neq 0$ cases.  Extending our analysis to a three-orbital model, we further identify an orbital-selective electron-hole pairing state.  These findings indicate that competition among multiple pairing channels in exciton insulator systems can significantly reduce  $T_c$ even in the presence of a large exciton binding energy. Our results illuminate the intricate interplay among Coulomb interactions, carrier density, and orbital degrees of freedom, providing valuable theoretical insights into the behavior of excitonic insulators in strongly correlated materials.

\textit{Acknowledgment} We thank Sheng Chen from the joint Ph.D. program of Sun Yat-sen University and Great Bay University, for useful discussions. This work is supported by the National Natural Science Foundation of China (Grants No.12274472, No.12494594). We also thank the support from the Research Center for Magnetoelectric Physics of Guangdong Province (Grants No.2024B0303390001) and Guangdong Provincial Quantum Science Strategic Initiative (Grant No.GDZX2401010). 

\bibliography{ref.bib}

\onecolumngrid

\end{document}